# Ultra-bright Coherent Undulator Radiation Driven by Dielectric Laser Accelerator

Yen-Chieh Huang[1,**] and Robert L. Byer[2,*]

[1]Institute of Photonics Technologies, Department of Electrical Engineering, National Tsing Hua University, Hsinchu 300044, Taiwan

[2]Department of Applied Physics, Stanford University, Stanford, California 94305, USA

*Corresponding author: Robert L. Byer, rlbyer@stanford.edu

**Co-corresponding author: Yen-Chieh Huang, ychuang@ee.nthu.edu.tw

**Abstract** – A dielectric laser accelerator, operating at optical frequencies and GHz pulse rate, is expected to produce attosec electron bunches with a moderate beam current at high energy. For relativistic electrons, the attosec bunch has a spatial length of a few nanometers, which is well suited for generating high-brightness superradiance in the VUV, EUV, and *x*-ray spectra. Our study shows that the brilliance of coherent undulator radiation driven by a short-bunch beam with 1~10 fC bunch charge from a dielectric laser accelerator is comparable to or higher than that of a synchrotron in the 0.1 ~ 3 keV photon energy range, even though the beam power of the dielectric laser accelerator is about a million times lower than that of a synchrotron. When the brilliance under comparison is normalized to the electron beam power, the proposed coherent undulator radiation source becomes the brightest source on earth across the whole VUV, EUV, and soft *x*-ray spectrum.

## I. Introduction

An electron accelerator is a powerful tool for both scientific investigations and industry applications. Recent studies on laser-driven particle acceleration [1-3] have shown a great potential toward applications driven by tabletop high-gradient particle accelerators. The electron beam from a particle accelerator has been known to be useful for radiation generation. For example, a synchrotron is built to generate spontaneous emission of electrons while circulating the electrons in a ring. The efficiency of electron radiation strongly depends on the ratio of electron bunch length to radiation wavelength. A short electron bunch can generate intense coherent radiation at a wavelength comparable to or longer than the bunch length. This intense radiation is dubbed as electron superradiance [4]. In a free-electron laser (FEL), the injected electron bunch from a conventional radio-frequency accelerator (RFA) is usually much longer than the radiation wavelength. However, the radiation feedback in an FEL structure can gradually modulate the electron density to form a micro-bunch train with a period comparable to the radiation wavelength. Coherent radiation from the micro-bunch train builds up an intense FEL output. Such a stimulated emission process takes place in both oscillator-type FEL and self-amplified-spontaneous-emission (SASE) FEL. The latter has been proven useful for *x*-ray laser generation without a resonant cavity. With its high brightness, a SASE *x*-ray FEL (XFEL) has earned the name as the 4$^{th}$ generation light source. Compared with the stimulated emission process in an FEL, electron superradiance is a much simpler coherent spontaneous emission process, except that one has to prepare bunched electrons to inject into a radiation generating structure.

The electron bunch length is usually a small fraction of the driving wavelength of an accelerator. In an RFA, an electron bunch length as short as ~0.1% of the driving wavelength [5] can be obtained through one of several bunching schemes, including photo-injection, velocity bunching [6], and magnetic compression [7]. In a laser-driven particle accelerator, the driving wavelength (and thus the electron bunch length) can be much shorter. For example, the electron bunch from a laser wake-field accelerator (LWFA) can be just a few femtoseconds [8], and recent experiments [9,10] have demonstrated attosec electron bunches from the so-called dielectric-laser accelerator [11,12] (DLA). For high-energy electrons, an attosec bunch could have a spatial length of a few nanometers, which is shorter than an extreme-ultraviolet (EUV) wavelength or comparable to a soft *x*-ray wavelength. Usually, a DLA operates under the laser damage of a dielectric [13], providing an acceleration gradient close to 1 GV/m for a laser pulse width of about 100 fs. Furthermore, when a properly phased laser field in a dielectric structure is polarized transversely along the electron axis, the structure could serve as a dielectric laser undulator (DLU) to wiggle the electrons. Given the high field gradient and integrable architecture of dielectrics, a DLA-driven DLU is promising for realizing ultra-compact superradiant sources in the EUV and soft *x*-ray radiation spectrum [14].

In this paper, we study coherent undulator radiation [15] (CUR) driven by the short electron bunches from a DLA. We further compare the brilliance of the CUR with that of existing and planned synchrotrons. Figure 1(a) shows a possible DLA-driven DLU based on dielectric gratings with properly phased and polarized laser field in the structures. When an electron is much slower than the speed of light, the strong phase slip between the slow electron and a fast optical cycle limits the acceleration length in an accelerator. To accelerate sub-relativistic electrons, one could employ a THz photoinjector to quickly ramp up the electron energy to the MeV level. It is easier to pack high-energy electrons with a high density in an optical cycle, because the space-charge force is relatively weak for relativistic electrons. A subsequent high-gradient DLA continues to accelerate the beam to high energy. Such a two-stage approach is advantageous in accelerating a large bunch of low-energy electrons in a THz cycle and distributing them at high energy to many optical cycles. Since the THz injector can adopt an optical THz source pumped by the same laser driving the DLA and DLU, the electron's macro-pulse structures of the THz injector [16] and the DLA will be the same and fully synchronized to the DLU field. To compare, Fig. 1(b) illustrates the vastly different dimensions between a centimeter long dielectric grating for future particle acceleration and a few-hundred-meter synchrotron for today's radiation generation. The grating is fabricated on silicon for THz particle acceleration and radiation, having a 220-μm period and 50-μm electron channel along its 2.2-cm length. The synchrotron is exemplified by the Taiwan Light Source (TLS) and Taiwan Photon Source (TPS) with circumferences of 120 and 518 m, respectively. However, as will be shown below, the calculated brilliance for a DLA-driven CUR source is surprisingly similar to that of a synchrotron for photon energies between 10 and a few keV. In the hard *x*-ray regime, although not a subject of this study, a DLA-driven SASE XFEL producing ~100 keV photons was proposed in the past [17].

Since a high-energy DLA is not yet available, we extrapolate the known performance of existing RFA to determine the parameter range of a future DLA. In general, the basic physics of linear particle acceleration is independent of the driving wavelength. Although some quantum phenomena [18-20] appear to be emerging from dielectric laser acceleration, it is premature to estimate their consequence and impact. As the physics of the beam manipulation elements used for an RFA beam are also applicable to a DLA, the conventional beam elements for manipulating the phase-space particle distribution for an RFA beam could also be conceived for a future DLA. One important consequence of this extrapolation is that the electron bunch length of an accelerator is roughly proportional to the driving wavelength. Unlike a typical RFA having a cylindrically symmetric structure, a DLA thus far has largely used planar structures conveniently fabricated via lithographic patterning. The planar DLA geometry is characterized by a rectangular accelerating channel that is very narrow (on the order of a fraction of the laser wavelength) in one transverse dimension but which can be arbitrarily large in the other. The bunch charge in a DLA can thus be increased by increasing the transverse electron beam size along the large dimension. This offers an opportunity to fill up more bunch charge in an optical cycle without increasing the longitudinal bunch length of the electrons.

We arrange the content of this paper as follows. In Sec. II, we first study spontaneous radiation with and without superradiance for DLA and RFA beams, respectively, in the same radiation structure. The study is based on separately considering two conditions. Under the first condition, we set the same bunch rate for both the DLA and RFA beams and then find the required DLA bunch charge needed to equalize the radiated spectral power. Under the second condition, we set the DLA and RFA beams to have equal beam power and then determine the enhancement factor of the spectral power generated from a DLA beam for different bunch charges. The first condition matches to the near-term development of a DLA as a high-pulse-rate and low-bunch-charge accelerator. The second condition requires an innovative DLA structure to accelerate a high bunch charge at a high bunch rate. In Sec. III, we derive the formulas to calculate the peak and average brilliances of a DLA-driven CUR source. Finally, in Sec. IV, we compare the brilliance of the envisaged DLA-driven CUR source with that of existing or planned RFA-based synchrotrons and XFELs. Sec. V provides some discussion to conclude this work.

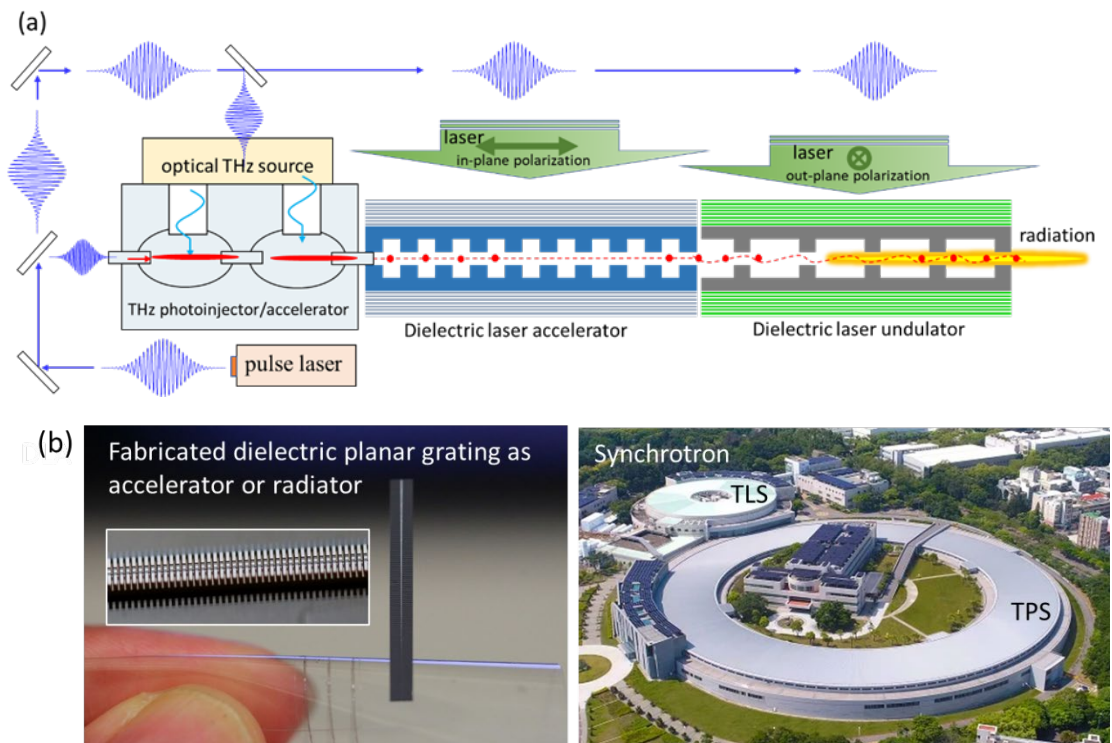

Figure 1. (a) A possible design of a DLA-driven CUR using dielectric gratings. Properly phased and polarized laser fields provide the electromagnetic forces to accelerate and wiggle the electrons. (b) Dimension comparison between a fabricated silicon dielectric-grating structure (left) held between plastic films on a fingertip and a synchrotron facility (right) exemplified by the 1.5-GeV Taiwan Light Source (TLS) next to the 3-GeV Taiwan Photon Source (TPS) with 120- and 518-m circumferences, respectively. Inset: zoom-in view of the dielectric grating with a 220-μm period and a 50-μm electron channel along the grating vector.

## II. Superradiance from DLA beam

In this study, we envision a DLA driven by a mode-locked infrared laser with sub-ps pulse length, 10s of µJ pulse energy, and a pulse rate of 1-100 MHz. Compatible laser parameters are now commercially available with solid-state lasers based on Ytterbium, Thulium, and Holmium [21]. Figure 2 shows a typical pulse structure, wherein every laser pulse accelerates a bunch train (or macro-bunch) of electrons containing $M_\mu$ micro-bunches. For a DLA driven by such a laser, $M_\mu$ is comparable to the number of optical cycles per laser pulse, which we take to be of order 100 (corresponding to a 330-fs laser pulse at 1 µm wavelength). If the macro-bunch repeats at $f_{laser}$ =10 MHz, the overall micro-bunch rate is $f_b = M_\mu f_{laser}$ = 1 GHz. This electron bunch rate is comparable or even higher than that of a typical synchrotron light source [22].

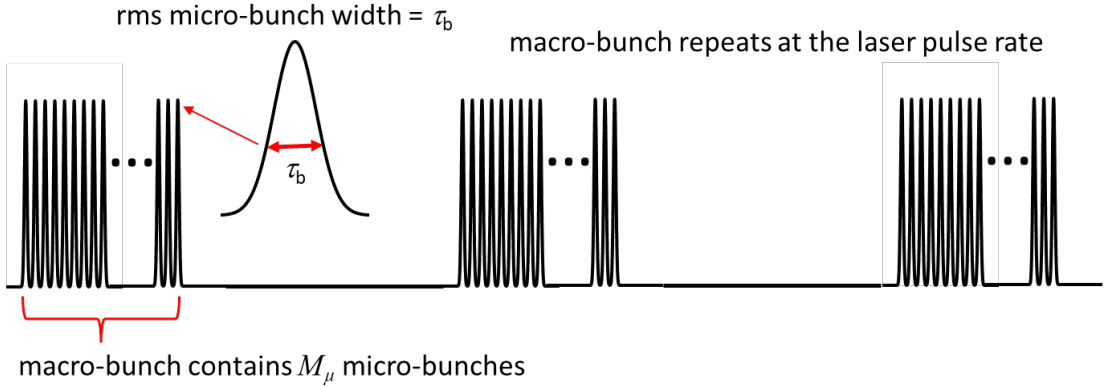

Figure 2. The electron bunch structure of a DLA. Each laser pulse drives an electron macro-bunch containing many micro-bunches matched to the optical cycles in the laser pulse. The effective bunch rate is equal to the laser pulse rate multiplying the number of micro-bunches in a macro-bunch.

Suppose there are $N$ electrons in each micro-bunch at the effective bunch rate $f_b$. Given a radiation device, the spectral power of the spontaneous radiation from the electron pulse train is expressed by [23]

$$P_N(\omega) = f_b W_1(\omega)[N(N-1)|b(\omega)|^2 + N], \qquad (1)$$

where $W_1(\omega)$ is the spectral energy radiated by a single electron at the angular frequency $\omega$, and $b(\omega)$ is the Fourier transform of the electron temporal distribution function $f(t)$ with the normalization $\int_{-\infty}^{\infty} f(t)dt = 1$. For a Gaussian electron bunch with a root-mean-square (rms) bunch length $\tau_b$, the normalized temporal distribution function is

$$f(t) = \frac{\exp(-t^2/2\tau_b^2)}{\sqrt{2\pi}\tau_b} \qquad (2)$$

and the modulus of its Fourier transform is

$$|b(\omega)| = \exp(-\omega^2 \tau_b^2 / 2) \leq 1. \tag{3}$$

The term $|b(\omega)|^2$ is the so-called electron bunching form factor, giving the magnitude of bunching at a particular radiation frequency. In the long-bunch limit $\tau_b \gg 1/\omega$, the bunching form factor vanishes and the spectral power in (1) is just linearly proportional to $N$. In the short-bunch limit $\tau_b \ll 1/\omega$, the bunching form factor is close to unity and the radiation power in Eq. (1) has a quadratic dependence on the electron number in a bunch or $N^2$. The short-bunch enhanced coherent radiation or superradiance can be intense when $N$ is large.

Wavelengths in the VUV, EUV and soft $x$-ray spectrum span from roughly 0.2 to 120 nm. The spectral range corresponds to 0.01 keV – 6 keV photon energy. For our purposes we take as an example a EUV wavelength of 10 nm for the desired radiation. The electron bunch from an RFA (with bunch duration in the fs and ps regime) gives nearly zero value for $|b(\omega)|^2$ in this part of spectrum; whereas, the attosec electron bunch from a DLA can have a form factor close to unity. In such a limit, for the same bunching rate $f_b$ and $N \gg 1$ for both the DLA and RFA beams, the ratio of the spectral power generated by the DLA beam to that generated by the RFA beam as obtained from Eq. (1) is approximated by

$$\frac{P_{N_{DLA}}(\omega)}{P_{N_{RFA}}(\omega)} \cong \frac{N_{DLA}^2 |b(\omega)|^2}{N_{RFA}}, \tag{4}$$

where $N_{DLA}$ and $N_{RFA}$ are the number of the bunched electrons in the DLA and RFA beams, respectively. The condition for the spectral power radiated from the DLA beam to exceed that from the RFA beam $P_{N_{DLA}}(\omega) \geq P_{N_{RFA}}(\omega)$ is therefore

$$N_{DLA} \geq \frac{\sqrt{N_{RFA}}}{|b(\omega)|}. \tag{5}$$

Assuming fully coherent radiation with $|b(\omega)|^2 \sim 1$ from a DLA beam and a nominal 1-nC bunch charge ($N_{RFA} = 6.25 \times 10^9$) in an RFA beam, one obtains from Eq. (5) $N_{DLA} \geq 8 \times 10^4$ electrons or 13-fC bunch charge for a DLA. In other words, in the EUV and soft $x$-ray spectrum, even though a DLA beam contains ~$10^5$ times less bunch charge compared with that of an RFA beam, it can radiate a spectral power as that generated by a nominal RFA beam with the same bunch rate. The 13 fC DLA bunch charge is also in the range of the optimal bunch charge studied for an energy-efficient DLA [24]. For the same beam voltage, the average beam power is proportional to the multiplication of the bunch rate and the bunch charge or $f_b N$. Under the equal-spectral-power condition, the ratio of the DLA beam power to the RFA beam power is therefore of order $N_{DLA}/N_{RFA} \sim 10^{-5}$. This greatly reduced beam power for a DLA illustrates the potential of a DLA-driven radiation device as a compact, efficient, and affordable laboratory instrument.

To increase the DLA beam current, accelerating a high-aspect-ratio elliptical beam [25,26] or an array of parallel beams [27,28] has been proposed in the past. In light of these possibilities, we now consider an alternative scenario in which a DLA delivers the same beam power as an RFA. For equivalent beam energies, we have that $f_{b,DLA} N_{DLA} = f_{b,RFA} N_{RFA}$, and so the ratio of the DLA spectral power to the RFA spectral power in the same radiation structure is given by

$$\frac{P_{N,DLA}(\omega)}{P_{N,RFA}(\omega)} = [(N_{DLA} - 1)|b(\omega)|^2 + 1]. \tag{6}$$

In writing (6), we have again assumed that the RFA bunch length is much longer than the radiation wavelength, as it is usually the case in the EUV and x-ray regime. The ratio in Eq. (6) can be considered as a spectral-power enhancement factor for a DLA beam over an RFA beam with the same beam power. Figure 3 plots the spectral power enhancement factor defined in Eq. (6) versus radiation photon energy between 0.1-4 keV for a DLA charge of 10 fC (blue curve) and 1 nC (red curve) in a 1-attosec rms (2.35-attosec FWHM) bunch length. It is seen that the short electron bunch from a DLA is highly advantageous for generating short-wavelength coherent radiation toward soft x-ray. The assumed 10 fC bunch charge is consistent with the near-term development of a DLA. The red curve can be considered as an ambitious goal when the DLA bunch charge could progressively approach that of an RFA in the future.

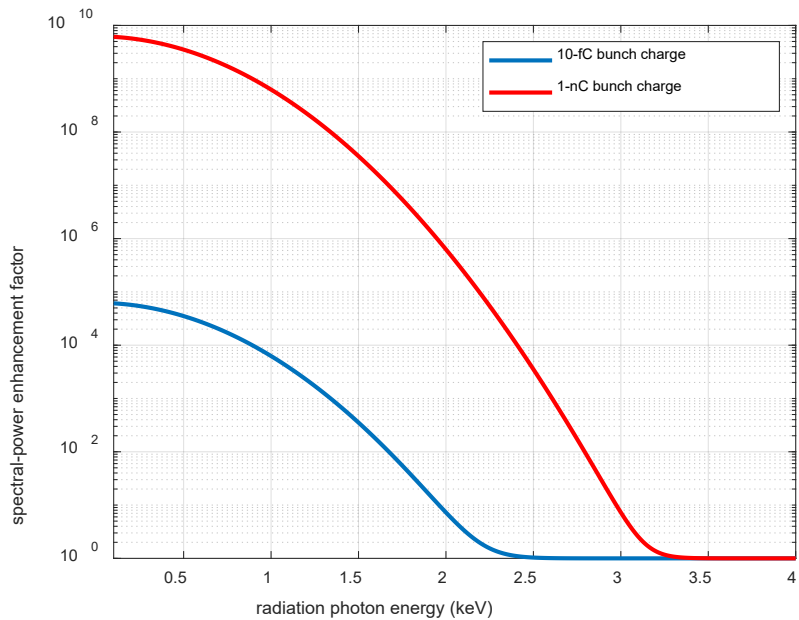

Figure 3. The ratio of the spectral power of a DLA-driven radiation source to that of an RFA-driven one vs. radiation photon energy, assuming 1- fC (blue curve) and 1-nC (red curve) charges in a 1-attosec long DLA bunch. In this calculation, the DLA and RFA have the same beam power. The ultra-short DLA bunch is highly advantageous in generating short-wavelength coherent radiation toward the soft *x*-ray spectrum.

### III. Short-bunch enhanced brilliance

The performance of a relativistic radiation source is usually described by its brilliance, which is the photon flux emitted per unit area, per unit solid angle in a 0.1% spectral bandwidth (BW). The commonly adopted unit is photons/s/mm$^2$/mrad$^2$/0.1%BW. The invariance source strength is the brilliance at the transverse phase space origin. For a single electron traversing an undulator, the brilliance at the phase space origin is given by [29]

$$B_1(\omega) = \frac{F_1(\omega)}{(2\pi)^2 (\sigma_r \sigma_{r'})^2}, \qquad (7)$$

where

$$F_1(\omega) = \frac{P_1(\omega)\Delta\omega}{\hbar\omega} \qquad (8)$$

is the photon flux with $\Delta\omega/\omega = 0.1\%$BW, $\hbar$ is the reduced Planck constant, and $\sigma_r$, $\sigma_{r'}$ are the rms radius and angle of the central radiation cone, respectively. By approximating the undulator radiation as a Gaussian laser beam, the radius and angle have the specific forms [29]

$$\sigma_r = \frac{\sqrt{2\lambda L_u}}{4\pi}, \sigma_{r'} = \sqrt{\frac{\lambda}{2L_u}}, \qquad (9)$$

respectively, where $L_u$ is the undulator length and $\lambda$ is the radiation wavelength. In this study, we focus on the most intense radiation at the fundamental wavelength along the undulator axis. At the phase space origin, the brilliance of a radiation source containing *N* radiating electrons in a bunch is written as [29]

$$B_N(\omega) = \frac{F_N(\omega)}{(2\pi)^2 \Sigma_x \Sigma_{x'} \Sigma_y \Sigma_{y'}}, \qquad (10)$$

where $F_N(\omega)$ is the photon flux with *N* electrons in $\Delta\omega/\omega = 0.1\%$BW, and $\Sigma_x \Sigma_y$ and $\Sigma_{x'} \Sigma_{y'}$ are the effective area and solid angle of the radiation cone, respectively. By including the coherent spectral power in Eq. (1), we modify Eq. (8) and write the photon flux from *N* bunched electrons as

$$F_N(\omega) = [N(N-1)|b(\omega)|^2 + N]F_1(\omega). \tag{11}$$

In the limit of small emittance, $\varepsilon_x, \varepsilon_y \ll \sigma_r \sigma_{r'} = \lambda/4\pi$, the phase space area of the radiation is approximately $\sqrt{\Sigma_x \Sigma_{x'} \Sigma_y \Sigma_{y'}} \sim \sigma_r \sigma_{r'}$ and the expression for the brilliance in Eq. (10) is recast into one for a diffraction-limited source with coherent radiation:

$$B_N(\omega) = [N(N-1)|b(\omega)|^2 + N]\frac{F_1(\omega)}{(\lambda/2)^2}. \tag{12}$$

For Eq. (12) to be valid in the EUV and soft x-ray regime, the condition for the transverse emittance $\varepsilon_x, \varepsilon_y \ll \sigma_r \sigma_{r'} = \lambda/4\pi$ has to be in the range of 1-10$^{-1}$ nm-rad.

The transverse emittance for most existing 3$^{rd}$-generation synchrotrons is in the range of nm-rad and that for diffraction-limited rings, such as the MAX IV [30], is already in the pm-rad range. With ~10$^5$-time reduced bunch charge in a DLA beam, the DLA beam emittance is likely to be smaller and it seems reasonable to assume the diffraction-limited brilliance in Eq. (12) for a DLA-driven radiation source.

For undulator radiation without short-bunch enhanced coherent radiation, the photon flux in 0.1%BW was previously calculated to be [29]

$$F(\omega) = 7.16 \times 10^{13} N_u [JJ] I \ (1/s) \tag{13}$$

where $N_u$ is the number of undulator periods, $I$ is the beam current in Amperes, and, for a planar undulator, the [JJ] factor is defined as

$$[JJ] \equiv 4A \times [J_0(A) - J_1(A)]^2 \tag{14}$$

with $A \equiv a_u^2/2(1+a_u^2)$. The parameter $a_u$ is the so-called undulator parameter, defined as $a_u = 0.093 \times B_{u\_rms}[kG] \times \lambda_u[cm]$ with $B_{u\_rms}$ being the rms undulator field.

To calculate the peak brilliance for a DLA-driven CUR source, we first write the current in Eq. (13) as $I = eN/\tau_s$, where $e$ is the electron charge and $\tau_s$ is the radiation pulse length. By using Eq. (1), we then replace $N$ with $N_{DLA}(N_{DLA}-1)|b(\omega)|^2 + N_{DLA}$ to include the short-bunch enhanced coherent radiation. For an electron bunch shorter than the radiation wavelength, the pulse width of the undulator radiation is simply the slippage length of the electron in the undulator [31] or $\tau_s = 2\pi N_u/\omega$. With these substitutions, Eq. (12) can be converted to an expression for the peak brilliance of a DLA-driven CUR source, given by

$$B_p(\omega) = 7.3 \times 10^{-6} \times [N_{DLA}(N_{DLA}-1)e^{-(\omega\tau_b)^2} + N_{DLA}]\frac{\omega[s^{-1}]}{\lambda^2[\mu m^2]}[JJ] \tag{15}$$

in units of photons/s/mm$^2$/mrad$^2$/0.1%BW. In the calculation, we have used Eq. (3) to include the bunching form factor of a Gaussian electron bunch.

For a given bunch rate $f_b$, it is straightforward to show that the average brilliance of a DLA-driven CUR source is given by

$$B_a(\omega) = \tau_s f_b B_p(\omega)$$
$$= 4.6 \times 10^{-5} \times [N_{DLA}(N_{DLA}-1)e^{-(\omega\tau_b)^2} + N_{DLA}] \frac{N_u f_b[s^{-1}]}{\lambda^2[\mu m^2]}[JJ]. \qquad (16)$$

The brilliance of a diffraction-limited light source is inversely proportional to the square of the radiation wavelength, as shown in Eqs. (15,16). However, the bunching form factor is vanishingly small when the radiation wavelength is shorter than the electron bunch length. As a result, the coherence-enhanced brilliance has a peak value near the electron bunch length. In the next section, Eqs. (15,16) are used to calculate the brilliance of an envisaged DLA-driven CUR source in comparison with the 3rd and 4th generation light sources. The interesting regime for comparison is where the radiation wavelength is longer than the electron bunch length of a DLA beam.

**IV. DLA-driven coherent undulator radiation**

In this section, we determine a set of design parameters for a DLA-driven CUR source optimized to radiate in the EUV and soft *x*-ray regime. The brilliance of this source is then compared with the 3rd and 4th generation light sources.

A DLA is driven by a laser field. For convenience, the drive-laser wavelength is fixed at 1 μm, which is $10^5$ times shorter than the 10-cm wavelength of an S-band RFA. Currently, a GeV RFA with a photocathode RF electron gun is capable for delivering ~ 1nC charge in a 100~200 fs bunch length [32]. Assuming that the electron bunch length is roughly scaled linearly with the driving wavelength, here, we adopt 2.35 attosec as the full-width-at-half-maximum (FWHM) bunch length (rms bunch length = 1 attosec) for a future DLA beam. From the bunching form factor defined in Eq. (3), the CUR cutoff photon energy for a DLA beam is approximately $\hbar\omega_c \sim \hbar/\tau_b$. For an attosec electron bunch, the cutoff photon energy is about 4 keV. We therefore limit our study in this paper to radiation photons below a few keV.

The amount of charge in an electron micro-bunch crucially influences the short-bunch coherent radiation. In Sec. II, under the same bunch rate, our calculation suggests a DLA bunch charge of 13 fC to achieve an average spectral radiation power equal to that from an RFA beam with 1-nC bunch charge. Furthermore, the optimal bunch charge for a photonic-band-gap fiber accelerator [24] was estimated to be ~5 fC. Therefore, for what follows, we choose 1 and 10 fC bunch charges for calculating the brilliance of DLA-driven CUR. The 1 and 10 fC bunch charges contain 6,250 and 62,500 electrons, respectively. Whenever there is a need for more charge to generate radiation, the characteristic planar structure of a lithographically fabricated DLA can accelerate more charge in an elliptical beam. The bunch charge for a DLA is therefore not strictly limited by its driving wavelength in the optical regime.

The electron bunch rate is equal to the multiplication of the laser pulse rate and the number of laser cycles in a laser pulse. As mentioned previously, a laser operating at ~10 MHz with tens of μJ pulse energy in a ~300-fs pulse width is readily available in the market. A DLA driven by such a laser will produce a 1-GHz electron bunch train.

A conventional magnetic undulator for synchrotron radiation is bulky and expensive. A DLU can be integrated with a DLA in a lithographic fabrication process. The undulator field of a DLU is derived from a properly phased and polarized laser field along the electron axis in a dielectric structure [17,33]. Like a DLA, a DLU operates under the laser damage field. The 1-GV/m peak field gives an equivalent peak undulator field of 3.3 Tesla. To generate EUV and soft *x*-ray radiations, the undulator period can be fixed at 1 mm for a DLA beam energy < 0.5 GeV ( $\gamma$ < 1,000). The undulator parameter $a_u$ in the [JJ] factor, Eq. (14), is therefore approximately 0.22. The undulator radiation bandwidth for the fundamental harmonic is the inverse of the number of undulator periods [34] or $\Delta\omega/\omega = 1/N_u$. To have the radiation in a 0.1% bandwidth, $N_u$ is chosen to be 1000.

In the ideal limit, the DLA operates at an acceleration gradient of ~1 GV/m to generate the assumed 0.5-GeV beam in a 0.5-m distance. The beam is injected into an assumed 1 m long DLU with a 1 mm long period to generate a radiation in a 0.1% bandwidth. The overall length of the proposed DLA-driven CUR source can be extremely compact, when compared with the size of an ordinary synchrotron with a circumference of a few hundred meters. For a 0.5-GeV beam containing 1~10-fC micro-bunches repeating at 1 GHz, the average beam power is only 0.5~5 kW. Table I summarizes the design parameters for the envisaged DLA-driven CUR source. To be practical, we shall compare some parameters listed in Table I with those published in the past for DLA research. For instance, the bunch charge and length of a 2 GeV DLA beam suggested in Ref. [17] for a DLA driven SASE FEL are 20 fC and 5 attosec, respectively, for a drive wavelength of 1 μm. The DLA bunch length adopted by Ref. [14] for radiation calculation is 1.7 attosec for the same 1-μm laser wavelength. In Ref. [35], the bunch charge, length, and repetition rate listed in Table VII for a 250 GeV DLA linear collider are 6 fC, 9.3 attosec, and 9.5 GHz, respectively, for a drive wavelength of 2 μm. For the equivalent undulator field in a DLU, Ref. [17] calculated a value of 4 T for a tilted-grating DLU and Ref. [35] adopts the same value of 3.3 T used in this Table I. Therefore, the parameters listed in Table I are fairly consistent with those in the literatures. In the following, we will use those parameters to calculate the brilliance of the proposed DLA-driven CUR in comparison with synchrotrons.

Table I Design parameters for an envisaged DLA-driven CUR source in the EUV and soft x-ray regime.

| System parameters | | | Remark |
|---|---|---|---|
| item | unit | quantity | |
| Driving laser wavelength, $\lambda$ | µm | 1 | 100,000th of the 10-cm S-band RF wavelength |
| FWHM bunch length, $\tau_b$ | attosec | 2.35 | scaled for the 1-$\mu m$ optical wavelength based on demonstrated 100~200-fs RF bunch |
| Bunch Charge | fC | 1, 10 | 6,250 and 62,500 electrons/bunch |
| Bunch rate, $f_b$ | GHz | 1 | 100 optical cycles in a ~300-fs pulse repeating at 10 MHz |
| Beam energy | GeV | < 0.5 | Used as a variable to tune the radiation wavelength |
| Undulator period | mm | 1 | a fixed value to radiate at $\lambda > 0.5$ nm for a beam energy < 0.5 GeV |
| Undulator parameter, $a_u$ | NA | 0.22 | 3.3-T peak undulator field under laser damage to dielectric undulator |
| Number of undulator periods, $N_u$ | NA | 1000 | radiation bandwidth ~0.1% |

By using Eq. (15), we show in Fig. 4 the calculated peak brilliance of the proposed DLA-driven CUR vs. photon energy with 1-fC (purple curve) and 10-fC (red curve) bunch charges. For comparison, we also show on the same plot the peak brilliance of the 3rd-generation accelerator-based light sources [36,37] and that of XFEL [38]. The DLA CUR brilliance first increases with the photon energy due to short-bunch enhanced coherent radiation and then the curve rolls off when the radiation wavelength becomes shorter than the bunch length. In the figure, the peak brilliance of the envisaged DLA-driven CUR is comparable or even higher than that of the 3rd generation light sources for photon energy below a few keV. This result is not surprising, because the peak DLA current used in our calculation is not much different from that of an RFA and yet the DLA's short bunch is highly efficient in generating EUV and soft x-ray superradiance. In Fig. 4, a DLA-driven CUR source cannot compete with an RFA-driven SASE FEL. However, it should be noted that the CUR calculation presented in Fig. 4 does not consider stimulated emission for the ultra-short electron bunches. On the other hand, when a DLA bunch length is significantly longer than a hard x-ray wavelength, a DLA-driven XFEL could be realized in the SASE regime with a comparable peak current[16]. We mark on the same plot (red diamond) the brilliance of such a SASE XFEL deduced from Ref. [17] for a radiation photon energy of 120 keV.

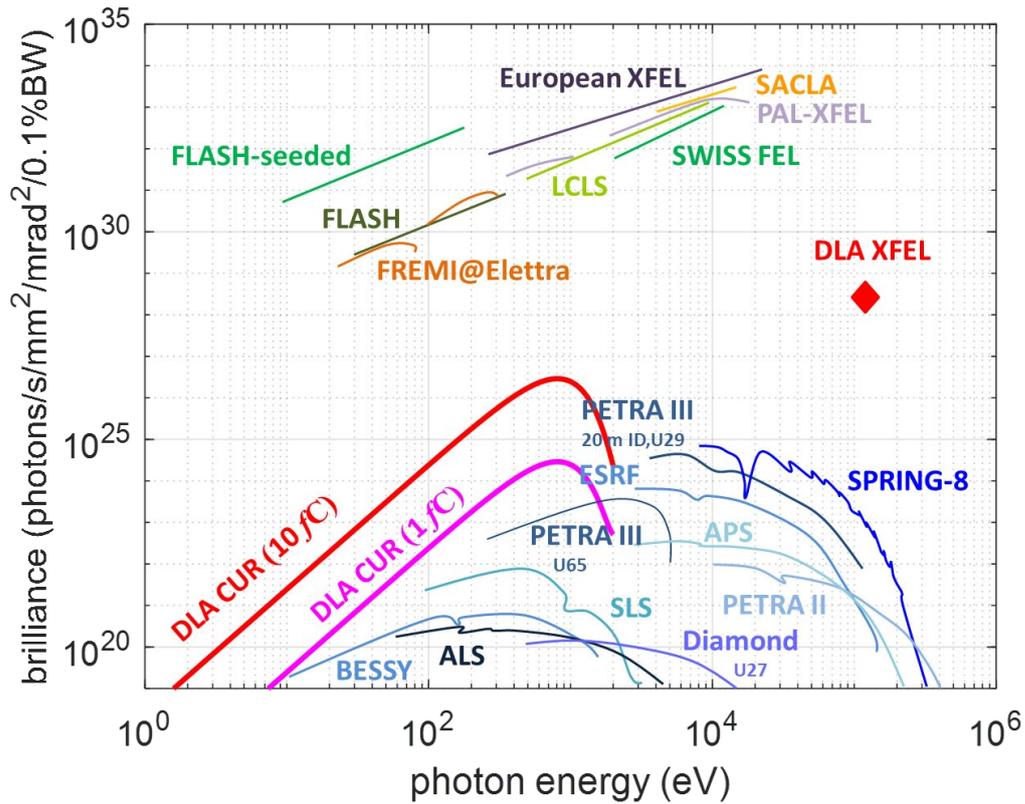

Figure 4. Peak brilliance of the DLA-driven CUR source (purple and red curves for 1- and 10-fC bunch charges, respectively) is superior to all known 3$^{rd}$ generation light sources for photon energy below a few keV, but is significantly lower than that of an RFA-driven XFEL. In the hard *x*-ray spectrum, the nano-bunch of a DLA beam generates less CUR, but can be useful for realizing an SASE XFEL in the hard *x*-ray spectrum [17] (red diamond).

Figure 5 shows the average brilliance of the DLA-driven CUR source versus photon energy calculated from Eq. (16) with 1-*f*C (purple curve) and 10-*f*C (red curve) bunch charges repeating at 1 GHz in comparison with that of existing and planned synchrotron rings. In the figure, the 1.5 GeV Taiwan Light Source [39] (TLS) is used to exemplify the existing rings below 2 GeV. The existing 2-8 GeV rings and new 3 GeV rings are approximately in the region denoted by purple and green bands [39], respectively. Marked in the overlapped green-purple band for comparison are the 3-GeV Taiwan Photon Source [39], 3-GeV Swedish MAX IV synchrotron [40], and 8-GeV Japanese Spring-8 synchrotron [39]. Dashed lines are estimated from planned upgrades and new constructions of synchrotron rings [36]. It can be seen from this comparison that, owing to the short-bunch enhanced coherent radiation and the 1-GHz bunch rate, the DLA-driven CUR source with 1-fC bunch charge has an average brilliance similar to the 1.5 GeV TLS. With 10-fC bunch charge, the DLA-driven CUR source is superior to existing 2-8 GeV rings and new 3-GeV rings in the 0.01-1 keV photon range. Again, the DLA CUR is not efficient in the hard *x*-ray regime due to the nanometer bunch length.

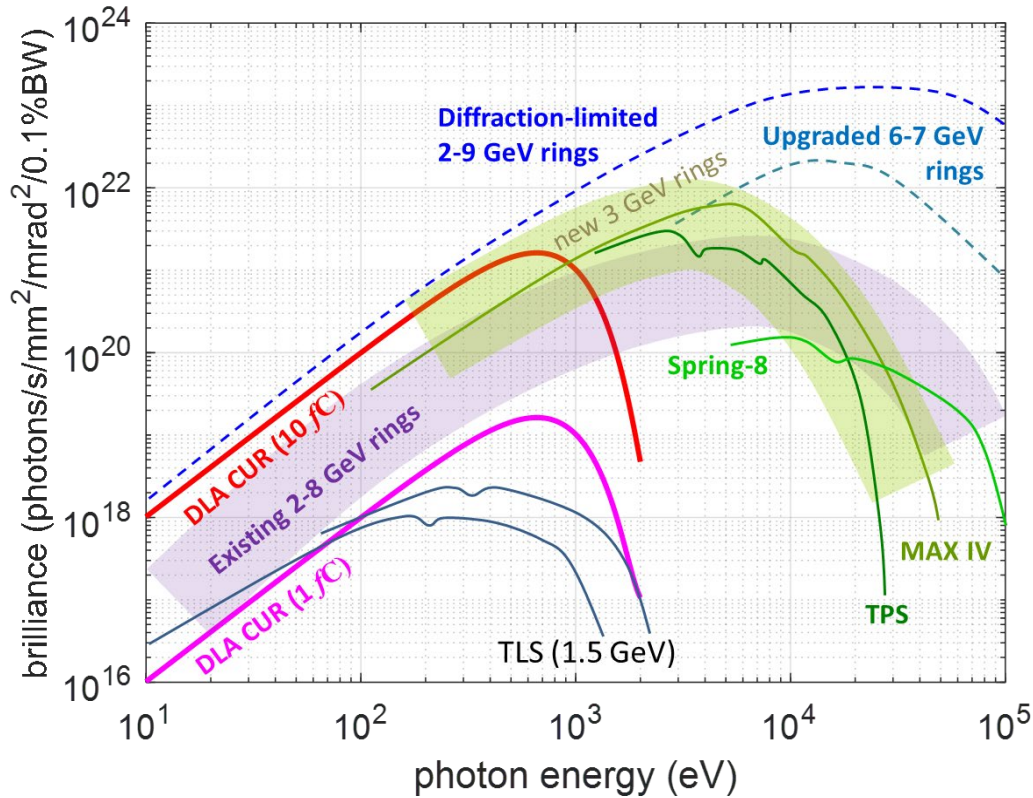

Figure 5. The average brilliance of the DLA CUR source (purple and red curves for 1 and 10 fC bunch charges, respectively) in comparison with existing and planned 3$^{rd}$-generation synchrotron sources. The DLA-driven CUR has comparable or higher average brilliance for photon energy less than a few keV.

With 1~10 *fC* bunch charge and 1 GHz bunch rate, the DLA beam current is only 1~10 µA. The average beam power for the 0.5-GeV DLA is in the range of 0.5 ~ 5 kW, which is not much different from the power consumption of a kitchen oven. However, for those new 3-GeV synchrotrons, the average beam current is usually between 300-500 mA and the circulating beam power is often higher than 1 GW. Given the 6 orders of magnitude difference in the beam powers, a fair and better comparison for the radiation sources can be the brilliance of a source normalized to the beam power. Figure 6 shows the average brilliance of the DLA-driven CUR and synchrotron rings normalized to the beam power in Watts. In the calculation, the DLA beam powers for 1 and 10 fC bunch charges are taken to be 0.5 and 5 kW, respectively. To illustrate the concept within an order-of-magnitude accuracy, we chose 1-GW beam power for all the synchrotron rings in this plot. Under such a comparison, the DLA-driven CUR source is clearly the brightest light source as a laboratory instrument across the whole VUV, EUV, and soft *x*-ray spectrum.

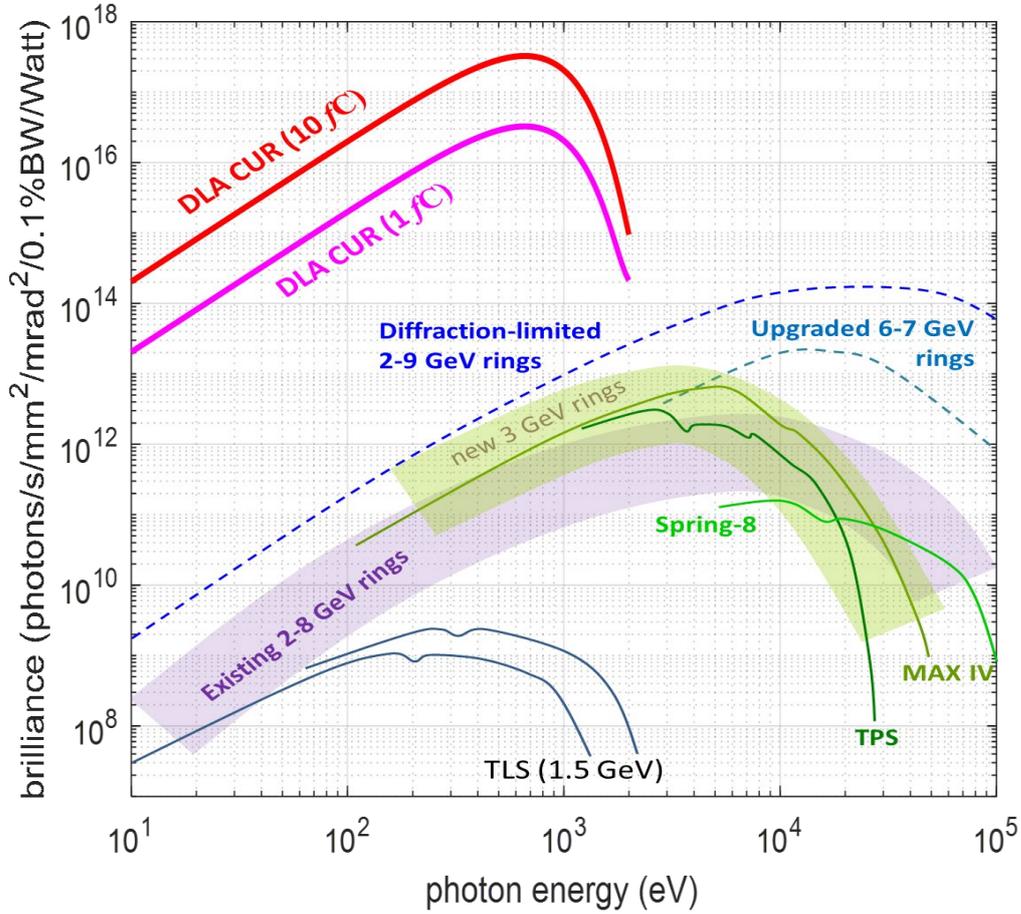

Figure 6. When normalized to the beam power, the average brilliance of a DLA-driven CUR source far exceeds that of a synchrotron source across the whole VUV, EUV, and soft *x*-ray spectrum. To illustrate the concept within an order-of-magnitude accuracy, the average beam power for all the synchrotrons is taken to be 1 GW in the calculation.

**V. Discussion and summary**

The approximately 5 orders of magnitude shorter drive wavelength for a DLA as compared with an RFA has several major advantages. The first advantage is the use of dielectric as the accelerator material to take a high laser field for high-gradient particle acceleration. The second advantage is the production of attosec electron bunches for efficient generation of short-wavelength radiation. When the electron bunch length is shorter than the radiation wavelength, the spectral energy of the radiation scales nonlinearly with the number of electrons in the bunch. We show in this paper that the attosec electron bunch from a DLA is well suited for generating high-brightness EUV and soft *x*-ray superradiance.

For the superradiance to be effective, the number of the electrons in a bunch must be nontrivial. In the EUV and soft *x*-ray regime, we determine in this work that, under the same bunch rate for DLA and RFA beams, an attosec DLA bunch with 13-fC charge can generate a superradiant spectral power comparable to that radiated by a nominal RF bunch with 1-nC charge. The corresponding difference in bunch charge by a factor of $10^5$ permits high brightness radiation from a low-power DLA beam. Furthermore, since a DLA is fabricated by lithographic patterning techniques on a planar wafer, the planar structure allows acceleration of more charge in a flat beam extended along one transverse direction. A multi-beam scheme for acceleration could further increase the bunch charge in a DLA.

For a given beam energy, the beam power is proportion to the beam current, which is the multiplication of the bunch charge and bunch rate of a beam. With increased bunch charge and bunch rate, the DLA beam power might progressively reach that of an RFA in the future. Figure 3 shows that, for a given beam power in a radiation structure, the average spectral power radiated by a DLA beam with 10-*f*C and 1-*n*C bunch charges can be millions to billions of times more intense than that radiated by a nominal RFA beam in the EUV and soft *x*-ray regime.

In this paper, the formula of the short-bunch enhanced brilliance is derived for a relativistic electron radiation source. By using the design parameters listed in Table 1 for DLA CUR, we show in Fig. 4 that the peak brilliance of the proposed DLA-driven CUR source is comparable to or higher than existing RFA-based 3[rd]-generation light sources for photon energy below a few keV. The superior performance of the DLA CUR is attributable to the comparable peak current and the short-bunch superradiance used in our calculation. In this comparison, the peak brilliance of the DLA-driven CUR source is still below that of FELs, because stimulated emission does not exist for an electron bunch with its bunch length shorter than the radiation wavelength. However, it has been shown previously that the high-brightness DLA beam has a potential to drive an ultra-compact SASE XFEL [17] in the hard *x*-ray regime.

Our study also shows that, with just 1-*f*C bunch charge, the average brilliance of the DLA CUR is already comparable to that of nominal synchrotron rings such as the 1.5 GeV TLS. With 10-fC bunch charge, the DLA-driven CUR can compete with existing, upgraded, and planned synchrotrons for photon energy between 10 eV and a few keV. It is interesting to note that the DLA beam power in our calculation is only 0.5~5 kW, whereas the typical beam power of a synchrotron is on the order of 1 GW. The proposed

DLA CUR source has a length of only a few meters from the accelerator to the undulator, whereas a synchrotron has a circumference of a few hundred meters. With the combined advantages of high brilliance, low beam power, and compact size, the proposed DLA CUR source can be a useful and affordable instrument in an ordinary laboratory. We further show in our calculation that, when the average brilliance is normalized to the beam power, the envisaged DLA-driven CUR source becomes the brightest radiation source across the whole VUV, EUV, and soft *x*-ray spectrum.

For both the DLA and laser-driven dielectric undulator, there are issues, such as injection, focusing, staging, power coupling etc., yet to be improved or solved for practical applications. Nevertheless, in the past two decades, the development of DLA has gained tremendous interest and momentum [41]. The conclusion of this paper could hopefully inspire the development of DLA as a useful tool to realize high-brightness and laboratory-scale radiation sources in the EUV and *x*-ray spectrum.

## IV. Acknowledgments and funding sources

Huang thanks helpful comments from R. J. England of SLAC. Huang is in debt to Ching-Shiang Hwang of NSRRC for the calculation of the average brilliance of TLS, TPS, and Spring-8 in Figs. 5 and 6. Huang is also thankful to Hossein Shirvani for his help on checking the calculations in this paper. This work is supported by the Ministry of Science and Technology, Taiwan, under Grants 110-2221-E007-103 and 108-2112-M-007-MY3.